\documentclass[times,twocolumn,final,authoryear]{elsarticle}

\usepackage{jasr}
\usepackage{framed,multirow}
\usepackage{amssymb}
\usepackage{amsmath}
\usepackage{latexsym}
\usepackage{graphicx}
\usepackage{epstopdf}
\usepackage{color}

\usepackage{amssymb}
\usepackage{bm}

 \def \pathdef  {fig/}
 \graphicspath{{\pathdef}}

\def\lsim{\;\raise0.3ex\hbox{$<$\kern-0.75em\raise-1.1ex\hbox{$\sim$}}\;}
\def\gsim{\;\raise0.3ex\hbox{$>$\kern-0.75em\raise-1.1ex\hbox{$\sim$}}\;}
\def\Msun{M_\odot}

\def\ergs{\rm ~erg~s^{-1}}

\def\ecsb{erg cm$^{-2}$ s$^{-1}$ arcsec$^{-2}$ }
\def\ecsb2{erg cm$^{-2}$ s$^{-1}$ arcsec$^{-2}$}

\def\lsi{LSI $+61^{\circ}\;303$ }
\def\psr2032{PSR J2032+4127}
\newcommand{\sla}{\;\raise0.55ex\hbox{\scriptsize$<$\kern-0.75em\raise-1.1ex\hbox{$\sim$}}\;}
\newcommand{\sga}{\;\raise0.55ex\hbox{\scriptsize$>$\kern-0.75em\raise-1.1ex\hbox{$\sim$}}\;}
\newcommand{\ssim}{\;\raise0.3ex\hbox{\tiny$\sim$}\,}
\newcommand{\sapprox}{\;\raise0.3ex\hbox{\tiny$\approx$}\,}

\def\apj{ApJ}
\def\apjl{ApJ Lett.}
\def\mnras{MNRAS}
\def\nat{Nature}

\def\prd{Phys. Rev. D}

\def\pre{Phys. Rev. E}
\def\araa{Ann. Rev. Astron. Astrophys.}                
 \def\aap{Astron. Astrophys.}                  
\def\aj{Astron. J.}                      
\def\apjs{Astrophys. J. Suppl. Ser.}                  
\def\apjl{ApJ Lett.}                   

\def\ssr{Space Sci. Rev.}
\def\jcap{J. Cosmol. Astropart. Phys.}

\def\aapr{ Astron. Astrophys. Rev.}

\usepackage{url}
\usepackage{xcolor}
\definecolor{newcolor}{rgb}{.8,.349,.1}

\usepackage[citebordercolor=white]{hyperref}

\journal{Advances in Space Research}

\begin{document}
\verso{A.M. Bykov \textit{etal}}

\begin{frontmatter}

\title{PeV proton acceleration in Gamma-ray Binaries}%

\author[1]{A.M. {Bykov}\corref{cor1}}
\cortext[cor1]{Corresponding author}
\ead{byk@astro.ioffe.ru}
\author[1]{A.E. {Petrov}\corref{cor1}}
\ead{a.e.petrov@mail.ioffe.ru}
\author[1,2]{G.A. {Ponomaryov}}
\author[1]{K.P. {Levenfish}}
\author[3,4]{M. {Falanga}}

\address[1]{Ioffe Institute, Saint-Petersburg, Polytechnicheskaya str., 26, 194021, Russia}
\address[2]{Peter the Great St.~Petersburg Polytechnic University, Saint-Petersburg, Polytechnicheskaya str., 29, 195251, Russia}
\address[3]{International Space Science Institute, Bern University, 3012 Bern, Switzerland}
\address[4]{Physikalisches Institut, University of Bern, Sidlerstrasse 5, 3012 Bern, Switzerland}

\received{}
\finalform{}
\accepted{}
\availableonline{}
\communicated{}

\begin{abstract} Current generation of ground based gamma-ray telescopes  observed dozens of sources of photons above 100 TeV. Supernova remnants, pulsar wind nebulae, young stellar clusters and superbubbles are considered as possible sites of PeV-regime particles producing the radiation.
Another possible source of PeV particles could be gamma-ray binary systems.
In these systems, a strong relativistic outflow from a compact object (neutron 
star or black hole) collides with the dense wind from a massive companion 
early-type star. Gamma-ray binaries
are observed from radio to high energy gamma-rays as luminous non-thermal sources. Apart from acceleration of very high energy leptons producing most of the non-thermal radiation, these systems may also efficiently accelerate protons. 
  We present here  the results of numerical simulation of the PeV-regime proton 
acceleration in gamma-ray binaries. The simulation is based on 
relativistic MHD modeling of local flows of magnetized plasma
in the region of interaction of two colliding winds.
We then inject 0.1 PeV protons into the system and directly follow their 
trajectories to demonstrate that they are accelerated to energies above PeV.
High magnetization of  the wind of the young massive star providing a Gauss range field in the winds interaction region is
of paramount importance for the acceleration of protons above PeV.
The maximum energies of 
protons accelerated by colliding winds in gamma ray binaries can significantly 
exceed the energy of the pulsar potential's drop, which limits
from above the energy of particles accelerated by an isolated pulsar.  
 \end{abstract}


\end{frontmatter}

\section{Introduction}
Identification of sources of PeV particles is of great interest for the problem of the origin of cosmic rays \citep[see, e.g.,][ and the references therein]{2019NatAs...3..561A,pevatron23,2018SSRv..214...41B,2021JPlPh..87a8401A,2021Univ....7..324C}.
Recently, the LHAASO observatory reported the discovery of 12 sources of ultrahigh-energy photons with energies up to 1.4 petaelectronvolts \citep{LHAASONat21}. The  source of the highest energy photon -- 1.4 PeV photon LHAASO J2032+4102 -- is located in the Cygnus region, which is very rich in potential cosmic ray accelerators.  
Real candidates for accelerators include the young massive stars Cyg OB1  and 
OB2, the microquasars Cyg X-1 and Cyg X-3, the supernova remnant $\gamma$-Cygni, 
the Cygnus Cocoon, and gamma-ray binaries
\citep[see e.g.][]{HEGRA,AckermannSB2011,2018ApJ...867L..19A,HonaNatAs2021,Tibet21}. 
One of the gamma-ray binaries --  PSR J2032+4127 -- is consistent in position with  
the source LHAASO J2032+4102. This indicates the need for theoretical 
modeling of gamma-ray binaries as potential sources of  protons in 
the multi-PeV range. The results of such modeling are presented in this paper.

Galactic magnetars and millisecond pulsars are considered for the
role of particle accelerators up to energies above PeV
 \citep[e.g.][]{arons03, arons12,2015JCAP...08..026K,lemoine15,Wilhelmi22}.
Recently,  \citet{proton_pulsa_cerutti20} performed a detailed 2D particle-in-cell modeling
of the acceleration of protons extracted from the surface of a neutron star-pulsar
with an axisymmetric magnetosphere. Their simulations took into account 
production of pairs and the magnetic field structure in the pulsar's magnetosphere, 
in which protons are accelerated in separatrix current layers inside 
the light cylinder of the pulsar. 
The highest energies, that protons in the model can reach, exceed $10\%$ 
of the total vacuum potential drop, and the luminosity can be several percent 
of the pulsar spin-down luminosity.
 General considerations have led \citet{Wilhelmi22}  to the following estimate 
  for the maximum energy $E_{\rm max}$  that particles (electrons 
 or protons) accelerated by an isolated rotating pulsar with spin-down power $\dot{E}$ 
 can receive:
\begin{equation}
E_{\rm max} \approx 2\, \eta_e\; \eta_B^{\, 0.5}\: \dot{E}_{36}^{\, 0.5}\;\; {\rm [PeV]}\, .
\label{epuls}
\end{equation}
Here  $\eta_e$ is the ratio of the electric and magnetic field strengths 
at the pulsar wind termination surface,  $\eta_B \leq$ 1 is the wind magnetization parameter, 
and $\dot{E}$ is measured in units of 10$^{36} \ergs$. 
 
The pulsars in binaries can accelerate particles to energies well above the limit given by Eq.\ref{epuls} in the region where the pulsar wind collides with the wind of 
the companion star. 
The  region where two MHD flows collide is known to be a favorable 
place for particle acceleration  \citep[see e.g.][]{EU93,Tavani_Arons97,B12,clumps_BR_13, deBecker13,BSPWN_2017, Grimaldo19,2020A&A...641A..84M, 2020MNRAS.494.3166V,PRV21,2023PhRvE.107b5201M,GR23}.
If the companion's wind is magnetized strongly 
enough to confine efficiently PeV particles in the system, those 
binaries could potentially accelerate particles to multi-PeV energies.

Such a system can be well represented by the gamma-ray binary 
PSR J2032+4127/ MT91$\:$213, which is considered as a possible identification for the PeV source LHAASO J2032+4102.
The binary is hosting a pulsar 
with a very eccentric orbit. The estimated orbital period is about $45$--$50$ years 
\citep[][]{Ho+17}. The pulsar has a spin period of about 143 ms, 
and an estimated spin-down power about $1.5\times 10^{35} \ergs$.
This spin-down power is almost an order of magnitude less than that would be 
required for an isolated pulsar to accelerate protons to PeV energies (see Eq. \ \ref{epuls}). 
In this regard, \citet{Wilhelmi22} pointed out that among the 11 very 
high-energy LHAASO sources which could be identified with some pulsars, 
only PSR J2032+4127 have a problem with such an identification when 
considering the particle acceleration mechanisms available to isolated 
pulsars.
However, in the gamma-ray binary PSR J2032+4127/ MT91$\:$213 particles accelerated in the pulsar vicinity can 
substantially increase their energy even further in the colliding winds flow. 
Monte Carlo simulation of colliding winds in the binary system 
PSR J2032+4127/MT91 213 under simplifying assumptions about their structure showed that for particles to reach PeV energies, a strong Gauss-range magnetic field 
in the wind of the massive companion-star is required \citep[][]{2032our}.
From the observational side, dipole fields exceeding 100 G are 
found in about $10\% $ 
of massive O, B, A type stars \citep{OBstars_Bfield17,Bstar_Bfield19}, 
and many of  Be type stars have magnetic field in the kG range.
Furthermore, \citet{Hubrig23} found that massive stars in 
binary and multiple systems (the progenitors of gamma-ray
binaries) are more likely to have strong magnetic fields.
The wind of such massive star in a binary 
should also carry a strong field.

In this work,  we are focusing on the very ability (or inability) of 
pulsars in gamma-ray binaries to accelerate ions above the limit given by 
Eq. \ \ref{epuls} and namely to energies 
of  $\sim 10$ PeV in PSR J2032+4127. To this end, we carried out a relativistic 
magnetohydrodynamical (RMHD) simulation of a pulsar wind nebula (PWN) which is interacting with the fast magnetized wind of a massive companion star, supplemented with a module (unit) for tracking the trajectories 
of accelerating ions (considered as test particles) inside the system.
Our simulation takes into account the effect of the Gauss-range magnetic field of the dense wind 
of a massive early-type star on the PWN's stucture, evolution and particle acceleration.
Such a field can greatly affect both the  acceleration and propagation 
of high-energy particles in the system,
allowing  gamma-ray binaries to become bright in the PeV range.

\section{Model description}\label{M}
Gamma-ray binaries consist of a massive early-type star 
and a compact object, a neutron star-pulsar or a black hole. A dozen of 
such gamma-ray loud objects, observed from radio to high-energy gamma-rays, 
are known for their high gamma-ray luminosity 
(\citealp[for a review, see, e.g.,][]{2013A&ARv..21...64D,Romero17}). 
Multi-wavelength observations of gamma-ray binaries at 
different orbital phases \citep[e.g., ][]{2018ApJ...867L..19A,2019ApJ...880..147N,2019ApJ...882...74H,2020MNRAS.497..648C}  
laid the  foundation for a numerical study of their structure and radiation \citep[e.g.,][]{takata17,2019A&A...627A..87C,2021A&A...646A..91H,2022MNRAS.511.1439F}. 

The binary systems with a
pulsar -- such as PSR J2032+4172, PSR B1256-63 and, 
likely, LSI $+61^{\circ}\;303$ -- power two colliding flows: the wind of a massive star
and  the relativistic wind of a pulsar.  In this work we are interested 
in systems in which a young massive star of early-spectral class O or B 
\citep[e.g.][]{Tavani_Arons97,Torres11} is orbited by
a rotation-powered pulsar. Rotation-powered pulsars 
are  known for their ability to inflate a fairly  energetic pulsar wind nebula. 
The interaction of  the nebula with the dense, highly magnetized 
wind of a massive star is of interest for the problem of cosmic ray acceleration.

There are three specific features which  distinguish a pulsar nebula  
in gamma-ray binaries from one created by an isolated pulsar. In the 
binaries, the nebular outflows are exposed to the intense radiation 
field of the young star. \citet{Ball_Kirk2000} studied whether  such radiation 
could slow down the pulsar wind 
(in the generic case of the gamma-ray binary PSR B1259-63) due to 
the inverse Compton drag force and found that the effect is insignificant. 
The second feature is the orbital motion of the pulsar in the binary. The motion 
can affect the structure of the region where the winds of the  
companions collide.
A thorough modeling of colliding winds in binaries was
carried out by \citet{Coriolis_Stone07} and 
\citet[][]{2012A&A...544A..59B} who studied the  effect 
of the Coriolis force on the PWN dynamics.
The third feature is the strong (Gauss-range) magnetic field of 
the wind of the massive star. Such a strong field has a profound
effect on the structure and dynamics of the nebular outflows and 
their very ability to accelerate particles above PeV energies, 
the latter being the main goal of our study.
Considering that the force of radiative braking is insignificant, 
let us evaluate the relative role of two other forces affecting the plasma dynamics in the colliding winds -- magnetic and Coriolis.
We consider the structure of the wind collision region 
in the rest frame of the pulsar. 
The pulsar moves with an angular velocity $\bm{\varOmega}$ along 
the Keplerian orbit.  In the pulsar's frame, 
the density of the Coriolis force applied to a plasma element of 
mass density $\rho$ is represented by the term $ -2 \rho\, 
\bm{\varOmega}\times\bm{u}$. In turn, the magnetic force density 
is given by the term  $[2(\bm{B} \, {\bf\nabla}) \bm{B} - {\bf \nabla} B^2]\, /8\pi$. 
The relative role of the forces can be 
evaluated through the parameter 
$\xi = 8 \pi \rho\, \varOmega R\, u/B^2$. 
Expressing the Keplerian frequency $\varOmega$ through the mass of the
young star $(M)$,  and the plasma density of the stellar wind  (the one 
of the pulsar wind is much lower) through the mass loss rate 
$(\dot{M})$, one can obtain
\begin{equation}\label{Cor}
\xi \;\approx \;0.05\; M_{10}^{\, 1/2} \:\dot{M}_{-7}^{{\color{white}{1/1}}} 
    R_{12}^{\, -1/2} \: B_2^{\, -2} \; ,
\end{equation}
if one assumes the equatorial supersonic stellar wind with the Parker-type 
magnetic field structure. In the latter case, the  magnetic field in the equatorial disk 
scales as $\sim R^{\,-1}$, so the surface magnetic field above $100\:$G would provide 
the stellar wind field $B_{sw}\sim \:$Gauss at the separation distance of $10^{14}\:$cm.
In turn, $\xi$  scales with the distance from  the young star as $R^{\,-1/2}$.
In the equation above, $M$ is measured in units of 10 $\Msun$, 
$\dot{M}_{-7}$ in $10^{-7} \Msun\,$yr$^{-1}$, $R_{12}$ in $10^{12}$ cm,  
and the dipole magnetic field $B_2$ at the stellar surface is in 100 G.

The relative role of the forces varies along the pulsar orbit. 
Estimates of the mass-loss rates of  O and early B stars (obtained from a sample of 67 stars)  
range from $\dot{M}_{-7}\sim$1 for O-type stars, to $\dot{M}_{-7} 
\sim$ $0.01$--$0.1$ for $\mbox{B0}$--$\mbox{B3}$ type stars \citep{massLOB19}. 
At the periastron distance ($\sim\:$AU) of the gamma-ray binaries 
known to host a pulsar, the influence of the Coriolis force may be relatively weak 
provided that the massive star's wind at this distance has a field in the Gauss range.
To estimate the extent of the structured magnetic disk of rapidly-rotating Be-stars  \citet{2021MNRAS.508.4887R} used MHD simulations and found  the disk extension to be several tens of the massive star radii that gives $\sim\:$ AU for typical radii values $\sim 10^{12}\:$ cm. Thus, at the periastron stage of the orbit high Gauss-range magnetic field provided by the equatorial disk may reach the vicinity of the pulsar. This allows to neglect the effect of the Coriolis force on the plasma flows structure in the colliding winds region and provides conditions for protons acceleration to multi-PeV energies.
Interestingly, the time to accelerate PeV proton in the PWN -- stellar wind collision region with the Gs-range field is a few hours. This sets the natural time period of simulation of PWN with the colliding winds structure to be about a day.
 
Our relativistic MHD model comprises three components:
(i) the stellar wind, (ii) the pulsar wind, and (iii) the test-particle relativistic cosmic rays interacting with the plasma flows of the colliding winds.
The model components used in numerical simulations 
are described in Appendixes. The choice of the model parameters (listed model-by-model in the legend to Table \ref{T1}) is explained below. 

(i) The massive star's wind.
Hot, rapidly rotating Be-stars may form an equatorial disk of rotating plasma 
\citep{Be_RiviniusAARv13}. An extent of such structured magnetic disk is difficult to 
infer from observations, but MHD simulation by \citet{2021MNRAS.508.4887R} gives it 
an estimate of several tens of stellar radii.
On the surface of Be-stars, the magnetic field can reach 
several  kiloGausses \citep{Bstar_Bfield19}.  Such a strong field 
can seriously influence the structure and dynamics of the radiatively driven stellar 
wind \citep{magn_disk_Be_Owocki22}, especially in the equatorial disk. The magnetization 
of the plasma there can be so strong that it will channel the flow into a field-aligned 
configuration.

In the case of 
the gamma-ray binary 
PSR J2032-4127 -- MT91 213 (see \citet{Ho+17} for its orbital parameters)
three years after the passage 
of periastron the pulsar in this system was observed at a separation distance of 
$R \sim 20\:$ AU from a massive  Be-star and could pass through its equatorial disk 
according to the model by \citet{Klement+17}.   
The magnetic field could be of several Gauss if it decreased in the equatorial disk as $\sim 1/R$ 
from the value of several kG on the surface of the Be-star.
So, we assumed that at this stage of the orbit the stellar wind had  
a magnetic field $B_{sw}$ of several Gauss, a flow velocity $v_{sw} \sim 300\:$km$\cdot$s$^{-1}$,
and a number density $n_{sw}\sim3 \cdot 10^4\:$cm$^{-3}$ in the vicinity of the pulsar orbit. The reference values taken above 
are consistent with the stellar wind models by \citet{Klement+17}
A wind with such parameters could be magnetically governed at distances of 
up to hundreds of stellar radii, as can be seen 
from a comparison  of its magnetic  and 
ram pressures, $B_{sw}^2/8\pi$ and $\rho_{sw}v^2_{sw}/2$.

Being interested in the orbital phases where PeV proton 
acceleration is expected we place a pulsar wind nebula to develop 
within a dense, highly magnetized, disk-like stellar outflow 
with a field-aligned configuration. 
 To simplify the numerical simulations over the time periods of a day duration the wind of massive companion star was approximated  asymptotically away from PWN as  a stationary, homogeneous flow of given 
 pressure  and density. The magnetic field and  bulk velocity are co-directional and 
 have a fixed direction relative to the pulsar rotation axis. 
Note that  in this case 
protons can boost their energy mostly in the vicinity of the pulsar wind nebula since the simulated stellar wind is asymptotically homogeneous. In more realistic case of a clumpy stellar wind PeV particle acceleration is even more efficient.   
The stellar wind velocity $v_{sw}$ is actually the velocity of 
 the pulsar relative to the stellar wind, because we are considering the system in the
 pulsar rest frame. The speed of the pulsar in orbit is 
 tens to hundreds km$\cdot$s$^{-1}$, so any value of $v_{sw}$ of this order 
 seems to be reasonable. The stellar wind model parameters are listed in Table \ref{T1}.
 There  $\psi$ represents the angle between the $\bm{B}_{sw}$ and the pulsar's rotation 
 axis. Here we take  $\psi=45^\circ$ to illustrate the effects and defer
 analysis of the consequences of this choice until a dedicated study.

 (ii) The pulsar wind model, the second component of our model system, 
 is described in details in Appendix B. The model was introduced 
 in \citet{Porth+14} and now is widely used in simulating pulsar wind 
 nebulae (for references see, e.g., \citealp{Porth17}). 
 As stated above we place our model pulsar wind nebula to develop 
 within a dense, highly magnetized  stellar wind. Under these conditions, 
 the nebula turns out to be much  more compact than it would be if it were 
 developing inside an evolved supernova remnant.  As a result, it quickly  develops its characteristic flow pattern. For the nebula, this takes no more than 
 a few hours, even if it is inflated from a scratch (which is the  case  
 in all our runs). Then high energy cosmic ray particles injected there 
 can  be effectively  upscattered and further accelerated by magnetic inhomogeneities  which are carried by the mildly relativistic flows in the vicinity of the pulsar wind termination shock.

(iii) Test-particle protons of PeV energies are the third component of our model 
system. We demonstrated below that the relativistic ions can increase their energy to $\sim$ 10 PeV in just a few successful head-on collision with  magnetic inhomogeneities of mildly-relativistic velocities. Scatterings on the fast moving magnetic inhomogeneities revealed in our simulations (see Fig.\ref{f4}) -- which have spatial scales of about PeV particles gyroradii and move with Lorentz-factors $\Gamma > 3$ -- are especially efficient.
Since the acceleration process occurs locally in the vicinity of PWN  and takes only $\sim 10^5$ sec, it is
practically independent of the system's pre-history. That is, PeV protons
care little about the structure that the nebula had in parts of the 
pulsar orbit other than where their acceleration began. PeV protons  also have little regard
to the symmetry of the system. For them, only the spatial scale of the acceleration site
and the spectrum of its magnetic inhomogeneities  are important.
Recall again, that in this study we are modeling the very ability of the 
pulsars with $\dot{E} > 10^{35} \ergs$ in binaries to accelerate particles to multi-PeV energies.

Since the acceleration of PeV particles is practically insensitive to the symmetry 
and  the structure of flows far away from the acceleration site, 
we simulated the system of two colliding winds in a planar 2D geometry. 
This choice is motivated by the excessively large computational resources 
that would be required for 3D modeling of   
propagation of particles in time-dependent MHD flows.
To trace the trajectories of accelerated particles, the plasma flow simulations must have a numerical grid that covers a
wind collision  region of tens of AU in size and simultaneously resolve the  
spatial scales on the order of sub-PeV particle gyroradii  ($\sim 0.02$ AU
for a 0.1 PeV proton in a field of $1\:$G).

Concluding the description of the model, we once again emphasize that 
the processes of interest to us here is the acceleration of multi-PeV particles. 
Both particle acceleration process and the adjustment of PWN's flows to the colliding stellar winds condition have similar time scales of the order of a few  hours. This is  short compared to the time of the orbital revolution of pulsars in most of the gamma-ray binaries of interest which are 
$\sim 26 \:$~days for LSI +61°303,  $\sim3.4\:$~yr for  PSR B1259-63, and $\sim 45$--$50\:$~yrs 
for PSR J2032-4127 \citep{Shannon+14, Ho+17}.

\section{Results}\label{R}

We conducted a series of numerical experiments with different stellar wind magnetic 
fields ($B_{sw}$) to study their influence on the ability of gamma-ray binaries 
to accelerate protons well above PeV.
In Figs. \ref{PWN_vs_B}--\ref{f5} we show the pattern of flows which is formed 
in a pulsar wind nebula  under the influence of a highly magnetized stellar wind.
The  pattern is  characteristic of a fully inflated nebula 
and develops just a few hours after the nebula begins to inflate, 
for any of 
$B_{sw}$ under consideration.
Again, this is much shorter  than the orbital revolution of a pulsar.
It seems to be a reasonable approximation  even for the system LS 5039 with the  shortest known orbital period of 3.9 days 
\citep{Casares05}. 
Therefore, this justifies considering the local structure of the wind collision 
system (i.e., considering a nebula in a short local part of the orbit) and fixing the directions of 
$\bm{B}_{sw}$ and $\bm{v}_{sw}$ on the simulation time scale, which is what we do in our simulations. 

\subsection{Local structure of the wind collision region}
\label{ssec:local_flows}

\begin{figure*}[h!]
\begin{minipage}{1.0\textwidth}
\begin{minipage}{.49\textwidth}
        \includegraphics[width=\textwidth]{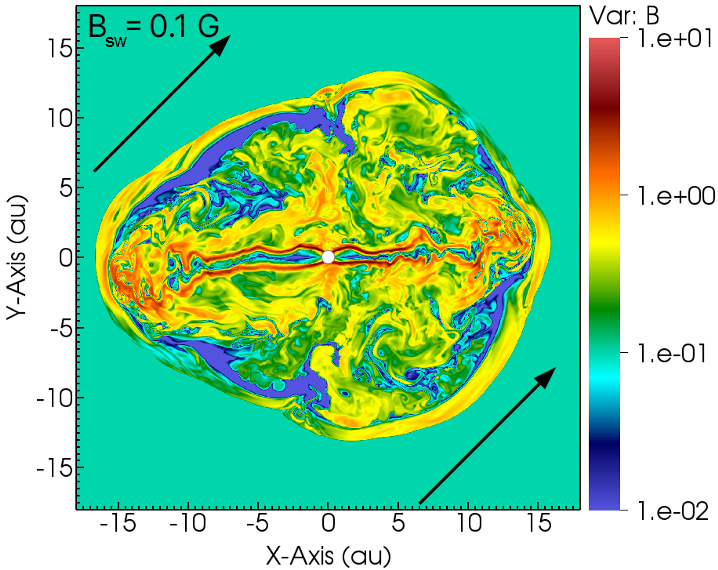} 
   \end{minipage}
   \nolinebreak
   \hfill
   \begin{minipage}{.49\textwidth}
        \includegraphics[width=\textwidth]{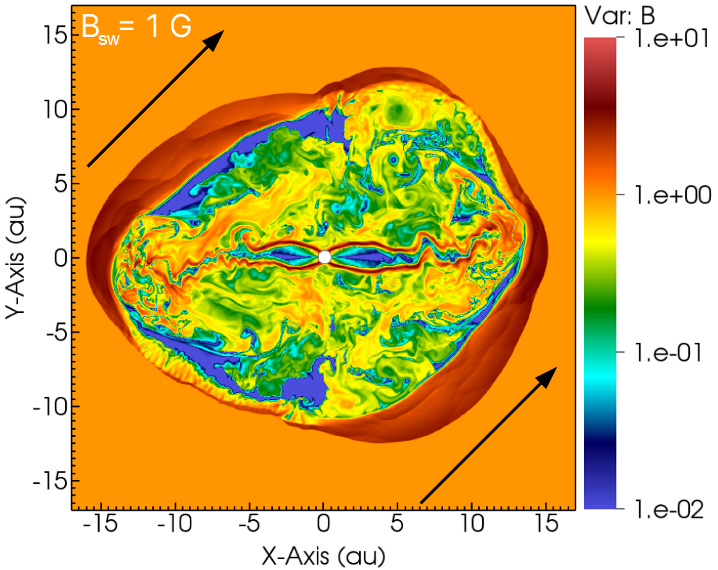}
   \end{minipage}
   \linebreak
   \vfill
   \begin{minipage}{.49\textwidth}
     \includegraphics[width=\textwidth]{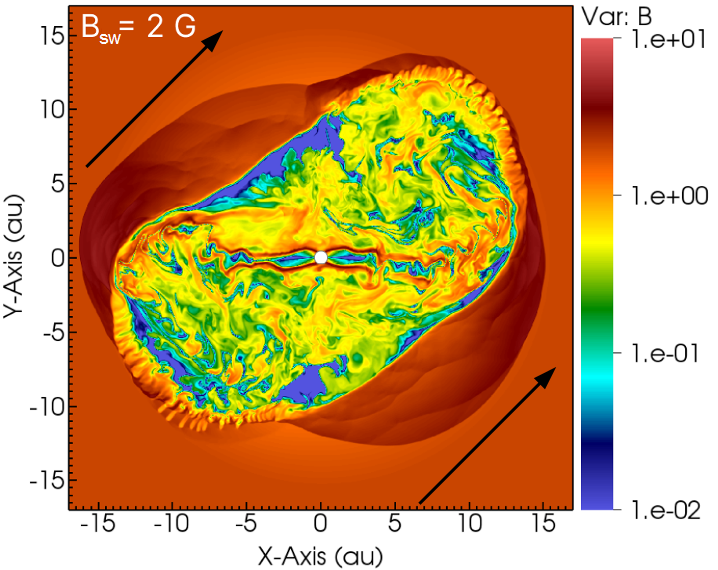}
   \end{minipage}
   \nolinebreak
   \hfill
   \begin{minipage}{.49\textwidth}
     \includegraphics[width=\textwidth]{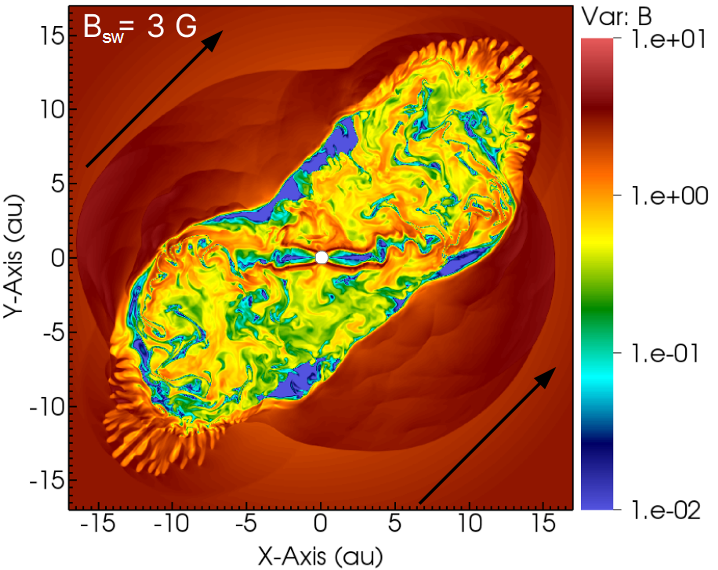}
   \end{minipage}
 \caption{Local structure of the region where two winds collide: the wind of a pulsar and the wind 
 of a massive star. Each panel of the figure shows a map of the local magnetic field $B$ [G] 
 in the region. The four panels differ in the strength of the stellar wind magnetic field:
 $B_{sw}=\,$0.1, 1, 2 and 3 Gauss (models $A$--$D$ in Table \ref{T1}). The white circles indicate the pulsar position, while the massive star is located outside the computational domain. The black arrows indicate 
 the direction of  $\bm{B}_{sw}$. It is oriented at an angle $\psi = 45^{\circ}$ 
 to the pulsar's rotation axis   (the latter is directed upward in the maps).
 The pulsar wind nebula in the figure is $t = 0.56\:$days old.
Note that the brightness scale of the color bar is adjusted to highlight the structure of MHD flows and does not reflect exactly the maximum and minimum values of $B$ in the region.}
 \label{PWN_vs_B}
  \end{minipage}
 \end{figure*}

Fig.\ \ref{PWN_vs_B} visualizes the local structure 
of the wind collision region and its change due to an increase in $B_{sw}$. 
Four panels in the figure illustrate cases with  $B_{sw}= 0.1$, $1$, $2$ and $3\:$G.
In the dense stellar wind, a PWN creates a cavity of characteristic dimensions 
of several AU. This is much smaller than what the same nebula would create if its 
pulsar were isolated and surrounded by the tenuous plasma of a supernova remnant.
In a smaller cavity, magnetic field energy accumulates much more rapidly. By the time 
the nebula is fully inflated, its  fields reach the sub-Gauss range 
(which is high compared to the tens or hundreds of $\mu$G for the fields of 
PWNe in supernovae).
Reducing the size of the nebula and enhancing its magnetic 
irregularities help increase the efficiency of accelerating high-energy 
particles in the system (Fig.\ \ref{part_SED}).

The stellar wind surrounding the nebula may have an even stronger field: 
 $B_{sw}\sim\:$Gauss. A strong field modifies the region of 
wind collision in several ways. Foremost, it most likely begins to control 
the structure and dynamics of the wind itself. 
As $B_{sw}$ increases, there is more and more reasons to expect that
the velocity field and magnetic field of the wind become co-directional.
Second, the strong  field reshapes the pulsar wind nebula, noticeably narrowing 
it across the field direction and slightly stretching along this direction 
(see Fig.\ \ref{PWN_vs_B}).  
When $B_{sw}$ reaches $ \sim 3\:$G, the nebula becomes nearly 3 times narrower  
in the transverse direction, than in the longitudinal.
At the same time, the total volume  
of the wind collision region
practically does not  change with increasing $B_{sw}$.
At sub-Gauss fields,  almost all of this volume is occupied by the PWN itself. 
As the field strengthens,
the nebula gives up more and more of this volume to a kind of dense cocoon that forms 
around it from the plasma of the massive star wind. Along most of the
perimeter of 
the nebula, the plasma in the cocoon turns out to be several times denser than  
the plasma outside the collision region. 
As a result, the field in the cocoon is  several times greater that the field 
outside. Even if the latter is in the sub-Gauss range, 
the cocoon field 
will still be $\sim\: $Gauss.
Clearly that wrapping the nebula in a  magnetic cocoon is of paramount importance 
for confining and accelerating particles of ultra-high energies within 
the region of wind collision.
This importance becomes even greater if we notice the development of 
magnetic inhomogeneities in the cocoon, which appear despite the fact 
that the wind outside the cocoon is assumed to be completely uniform in our runs.

Fig.\ \ref{f4} characterizes the collision region from a slightly different aspect
(model C in Table \ref{T1}). 
There we show a map of flow velocity,  in units of $c$   in the upper panel
and in units of the Lorentz factor $\Gamma$ in the lower panel.
The map shows signs that in the windward side of the region, mildly relativistic flows 
may occupy a larger part of the available volume than in the leeward side. In the lower panel
we zoom in on the surroundings of the wind termination shock. There, for clarity, 
a contour plot of the Lorentz factors is superimposed on a gray-color map of 
magnetic field magnitudes. The panel demonstrates the  presence in the system 
of mildly-relativistic magnetized  flows  with  $\Gamma \lsim 3$, as well as individual
magnetic clumps with even higher $\Gamma \sim 4.5$ and with characteristic sizes 
comparable to the gyroradii of PeV-regime particles.
 
\subsection{High energy particle acceleration}\label{PA}
Proton gyroradii scale with magnetic field strength as $R_g \approx 3 \times 10^{12} 
\cdot E\cdot B^{-1}$\:cm, with  proton energy $E$ measured in PeV and $B$ in Gauss. 
Accordingly, fields in the Gauss-range 
are necessary to confine particles with energies of several PeV in the collision region of several
AU in size.  Then the acceleration time and the spectral energy distribution of the test particles 
will be determined by the broad-band spectrum of magnetic fluctuations in the acceleration region. 
Due to the limited numerical resolution of our experiments  (aimed at simulating the trajectories
of the PeV-regime particles),  we cannot trace  these fluctuations down 
to scales resonant for GeV-regime particles. Therefore, we cannot consider fluctuations 
over a range of scales which is wide enough for kinetic or Monte Carlo modeling
of the spectral energy distribution 
over many decades
\citep[see e,g,][]{BE87,BSPWN_2017}.
In this regard, we will limit ourselves to studies of the propagation and acceleration of test-particle 
protons of PeV energies.

\begin{figure}[h!]
\includegraphics[width=\columnwidth]{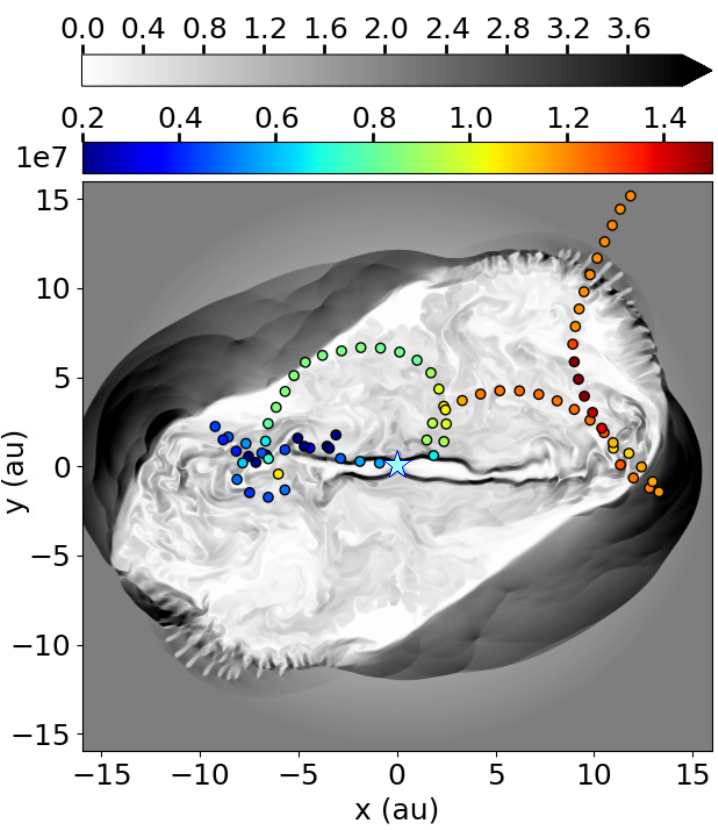}
\caption{
     Trajectory and energy gain of an accelerating proton 
     with an initial energy $2.4 \times 10^{15}\:$eV. 
     The test-particle proton were
     injected near the pulsar wind termination shock into the wind collision region.
     The trajectory is traced by thick dots, and the gain (in terms of the Lorentz factor) 
     is shown by the color of the dots.
The trajectory is modeled using a time-dependent model, but  for clarity 
it is superimposed on a still map of the system's magnetic field $B$ 
(in G; top gray color bar); the blue star indicates the pulsar position.  
As can be seen, the proton left the system, gaining energy above 
10$^{16}$ eV, after several scatterings in a mildly relativistic MHD flow. 
The system of colliding winds is simulated for a  stellar wind magnetic field 
of $2\:$G and a pulsar spin-down luminosity of $10^{37}\:$erg$\cdot$s$^{-1}$ (model C in Table \ref{T1}).
The brightness scale of the map is adjusted to highlight 
the structure of RMHD outflows and  does not reflect the maximum and minimum
values of $B$ in the system. The map shows the system  $0.56\:$ days old since 
its inflation began.
}
\label{f5}
\end{figure}

A test-particle proton injected into the region of two colliding winds is shown 
in Fig.\ \ref{f5}. The figure visualizes how this proton changes its energy as it 
wanders within the region.  To guide the eye, the proton's trajectory is superimposed 
on a still map of the region's magnetic fields. Note that the choice of initial energy 
for an injected  particle is limited both from below and from above. From below, it is bounded 
by the adopted resolution of the numerical grid. However, if lower energy (TeV-regime) 
particles were injected into our system,  they would be successfully confined and 
accelerated,  although much higher resolution runs are needed to allow RMHD flows 
to effectively scatter such particles. From above, the initial energy is bounded  
by the maximum particle energy available in the source of injection.
This source can be associated either with the colliding winds region or with the termination 
surface of the pulsar wind (\citealp[see, e.g., ][]{amato_arons06}). In the later case the initial 
particle energy is limited  by the drop in electric potential between the pulsar and infinity 
(\citealp[for a review, see][]{arons12}). According to a recent detailed discussion of particle 
acceleration by young bright pulsars in \citet{Wilhelmi22}, for protons to reach a PeV energy, 
a pulsar spin-down luminosity about 10$^{36} \ergs$ is required. In gamma-ray binaries, 
the energy of injected proton can be further boosted in the region of winds' collision.

\begin{figure}[h!]
\begin{minipage}{1.0\columnwidth}
\center{\includegraphics[width=1.0\columnwidth]{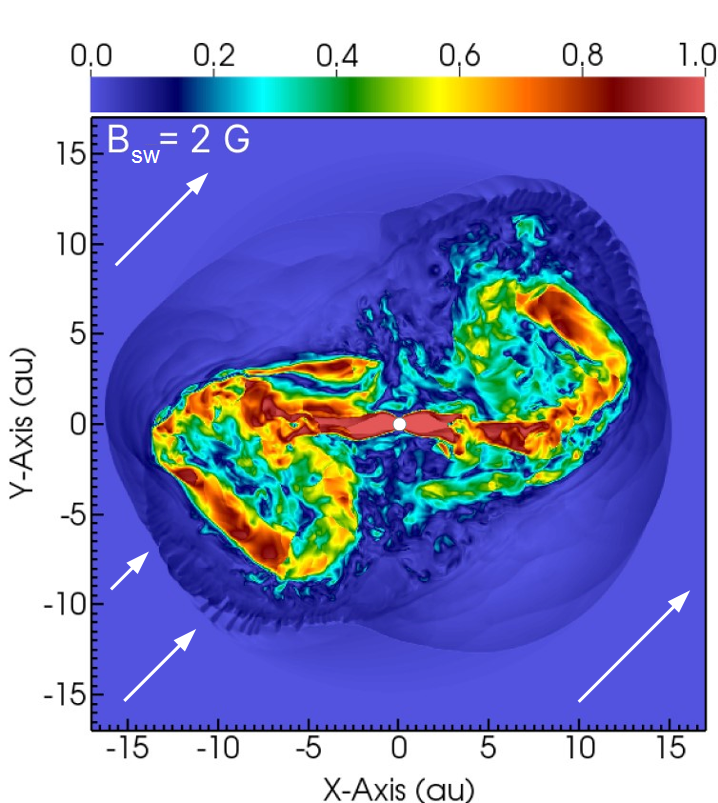}}
\end{minipage}
\linebreak
\vfill
\begin{minipage}{1.0\columnwidth}
\includegraphics[width=1.0\columnwidth]{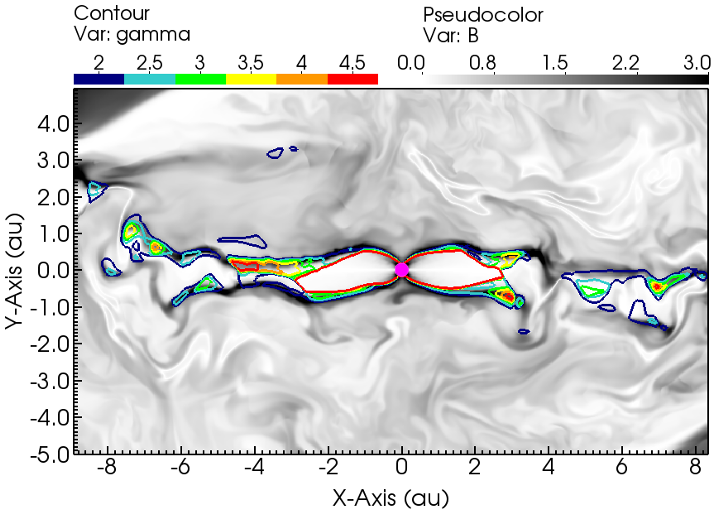}
\end{minipage}
\caption{
Local structure of the wind collision region. A model with
a stellar wind field of $2\:$G and a pulsar spin-down luminosity
 of $10^{37}\:$erg$\cdot$ s$^{-1}$ is presented (model C from Table \ref{T1}).
Top: Map of flow velocity magnitude (in units of the light speed $c$). White arrows indicate the direction of the stellar wind flow.
Note that on the windward side of the nebula, high-velocity outflows 
occupy more of the available volume than on the leeward side.  
Bottom: Contour map of the Lorentz factor of flows, $\Gamma$. 
The map is superimposed on a  gray-color map of the local magnetic field $B$ (in G). 
The  $\Gamma=4.5$ contour (shown in red), outlining the region of the 
pulsar's unshocked wind, 
practically coincides in position with     
the wind termination shock due to an abrupt deceleration of the wind.
The brightness scale of the gray colorbar is adjusted to highlight 
the structure of RMHD outflows and does not reflect the maximum 
and minimum values of $B$.
The nebula in the figure is $0.56$ days old from the start of inflation. The white circle in the top panel and the pink circle in the bottom panel indicate the pulsar position.
}\label{f4}
\end{figure}

In Fig.\  \ref{f5}  thick circles visualize the proton's trajectory, and their color 
represents the proton's energy.
It can be seen that the particle reached about 14 PeV when it left the acceleration site. 
In our simulations, we traced trajectories of many thousands of protons, injected 
with different initial energy at different locations. The figure just illustrates 
the system's ability to confine multi-PeV protons and accelerate them above 10 PeV. 
This acceleration typically occurs 
in literally $2$--$3$ successful upscatterings
on  fast-moving 
magnetic inhomogeneities with the Lorentz-factors  $\Gamma \gsim 3$ (see in Fig.\ \ref{f4}).
Indeed, a head-on collision with the plasma inhomogeneity moving with Lorentz-factor $\Gamma$ may boost the particle energy in $\sim \Gamma^2$ times, like it happens with the photons undergoing the inverse Compton scattering.  Such mildly relativistic plasma structures are present in the downstream of oblique parts of the pulsar wind termination shock. 

\begin{figure}[h!]
\begin{minipage}{1.0\columnwidth}
\center{\includegraphics[width=0.98\columnwidth]{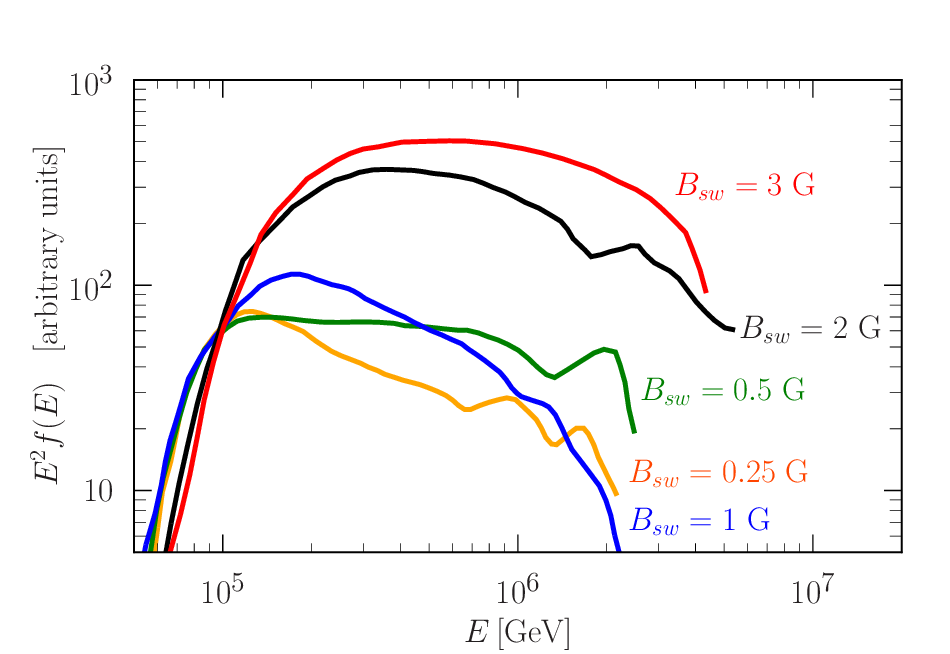}}
\end{minipage}
\linebreak
\vfill
\begin{minipage}{1.0\columnwidth}
\center{\includegraphics[width=0.98\columnwidth]{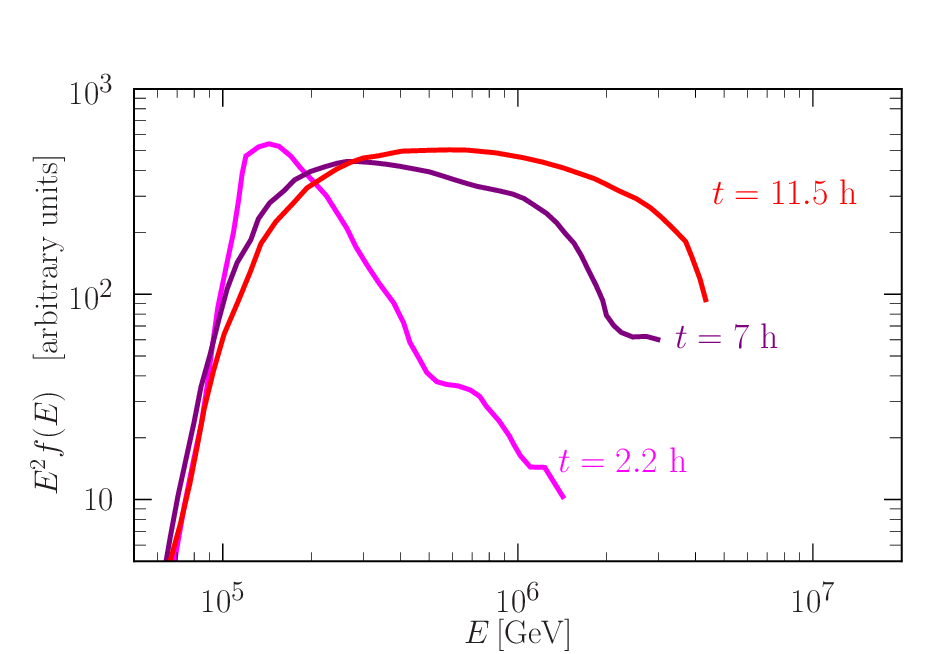}}
\end{minipage}
\caption{Spectral energy distributions of PeV-regime protons accelerating in 
a gamma-ray binary system, in models from \textsl{E} to \textsl{I} in Table \ref{T1}. 
Shown are the spectra of particles inside the simulation domain.
Top graph: Spectra of accelerating particles at different strengths of
the stellar wind magnetic field $B_{sw}$ (in Gauss units):
$0.25$ -- orange curve, $0.5$ -- green, $1$ -- blue, $2$ -- black, $3$ -- red.
In the illustration,  spectra were smoothed and taken in a system that had evolved 
from scratch for $18.5\:$hours, and $11.5\:$h after the particles were injected 
into the system.
 Bottom graph: Spectra of accelerating particles at
 different moments in the evolution of the system. In the illustration,
 we show the moments  $t= 2.2$ (magenta), $7$ (purple) and $11.5$ (red) 
 hours after  the injection of test particles for model \textit{I}, $B_{sw} = 3$\:G.
}
\label{part_SED}
\end{figure}

Fig. \ref{part_SED} shows the spectral energy distributions of accelerated particles 
obtained in models ``E" to ``I" in Table \ref{T1}. 
The upper panel illustrates how the stellar wind field $B_{sw}$ affects the particle 
spectra (formed some time $t$ after the particles were injected into the system).
As mentioned in \S\ref{ssec:local_flows}, high-magnitude fields $B_{sw}$ help 
accelerating particles stay longer in the acceleration region and encounter stronger 
magnetic irregularities there. All this promotes acceleration.
As $B_{sw}$ increases, particle spectra become harder, and the maximum energy achieved  
by accelerating protons grows. 
The lower panel illustrates how particle spectra evolve with time. 
It can be seen that particles injected into the system with an initial energy of 0.1 PeV
are quickly accelerated (in a few hours) to PeV energies. On the simulation timescale, 
particle spectra become increasingly hard.
    
\section{Discussion}
Currently, among the known gamma-ray binaries, there are two systems 
that definitely contain a pulsar, and one that is highly likely to contain.
The pulsar  PSR J2032+4127 orbits the  massive companion Be-star MT91 213 (B0Vp) 
in about $45$--$50$ years.  At periastron, the pulsar 
passes at a distance of $\sim 10^{13}$\:cm  from the Be-star. 
This is close enough for the pulsar wind nebula to find itself in 
a stellar wind magnetic field  of $\sim\:$Gauss magnitude, provided 
the star has a surface magnetic field  $\gsim100\:$Gauss.
A rapidly rotating Be-star may also form an equatorial disk of rotating plasma 
\citep{Be_RiviniusAARv13}. The  extent of such a disk is still difficult 
to constrain observationally. Using MHD modeling, \citet{2021MNRAS.508.4887R} 
estimated the extent of  the structured magnetic disk  to be  several tens of the massive star radii. 
If this is indeed the case, the  gamma-ray binary PSR J2032+4127   
may well be  a plausible counterpart to the LHAASO  source J2032+4102.
The other two gamma-ray binaries hosting a pulsar have much shorter 
orbital periods. In the first one, 
the pulsar PSR$\:$B1259-63 with a spin-down luminosity of 
$8\times 10^{35}\:$ erg$\cdot$s$^{-1}$ orbits the (apparently) Be star 
LS 2883 with a disk every 3.4 years 
(see e.g. \citealp[][]{2019A&A...627A..87C,2020MNRAS.495..365C}).
The periastron and apastron distances of the pulsar's orbit are 1 AU and 13.4 AU, 
respectively \citep{2013A&ARv..21...64D}.
In the second system, the gamma-ray binary LSI $+61^{\circ}\;303$,
a compact object  \citep[possibly a pulsar;][]{lsi_pulsar_nat22} orbits B0 Ve star 
in about 26.5 days. Its apastron distance is estimated to be $\sim 1\:$AU.
This binary exhibited a strong flaring activity in TeV-range photons
lasting about a day around the apastron position \citep{Lsi_TeV_flare16}. 
Clearly, in the three gamma-ray binaries mentioned above, pulsar wind nebulae 
have to evolve in a markedly different environment compared to that can be 
observed around the nebulae of isolated pulsars in supernova remnants or in 
the interstellar medium \citep[see e.g.][]{GS_06ARA&A, Reyn_SSRv17}.
The region of two colliding  MHD flows in these systems provides favorable conditions
for efficient acceleration of particles to energies well above PeV and their more 
intense high-energy radiation, which distinguishes gamma-ray binaries from the broader 
population of X-ray binaries.  This may resolve the problem of the association of the 
gamma-ray binary PSR J2032+4127/MT91 213 with the 1.4 PeV photon source LHAASO
J2032+4102.

Using the RMHD code PLUTO \citep{Mignone+07,Mignone18} we modeled 
the wind collision region in which PeV protons can boost their energy.
Our simulations revealed 
the character of plasma flows,
including their magnetic and velocity fields, in the vicinity of the pulsar 
on a spatial scale smaller than the distance between the stars.
The resolution of the applied numerical grid was suitable for studying the confinement 
and acceleration of very high energy particles of PeV regime. 
At pulsar spin-down rates of  $10^{35}$--$10^{37}\:$erg$\cdot$s$^{-1}$  
and  Gauss-range stellar wind fields,
a typical pulsar wind nebula acquires an elongated shape. It turns out 
to be $\sim 10\:$AU in extent in the field direction and noticeably smaller across.
This is roughly 3 orders of magnitude smaller than what the 
corresponding nebula would be if its parent pulsar were propagating at subsonic 
speeds in a supernova remnant. Because of their compactness, PWNe in gamma-ray 
binaries acquire characteristic flow patterns and accumulate energy much more rapidly 
than their remnant counterparts. They easily adjust their flow pattern 
to local conditions in the stellar wind anywhere in the pulsar's orbit, especially 
when the pulsar is crossing (or traveling in) the dense, highly-magnetized equatorial 
disk of a massive star.
Actually,  the evolution of MHD flows in PWNe and the acceleration of PeV protons
in these flows  occur on comparable time scales 
(from a few, up to several hours).
This allows one to limit the modeling of PeV protons acceleration 
to a short local part of the orbit,  which is essentially what we do:  
we consider the  \textit{local} structure of a wind collision region 
on time scales of the order of $\sim 10^5$ sec. This also suggests, as 
a first approximation, that the pulsar nebula sees the surrounding 
stellar wind as a plane, homogeneous flow. Moreover, if this flow 
carries a magnetic field of the Gauss-range, then it will most likely to be controlled by 
this field, so that the field of velocity and the magnetic field in this flow 
can be considered to a fairly good approximation as co-directional.

The accelerated particles could reach higher energies if a more complex 
model of the clumpy wind of a massive star  were  considered.   The presence of clumps 
in a stellar wind in gamma-ray binary systems hosting a Be-star and a pulsar is consistent
with multiwavelength observations 
\citep[see e.g.][]{owocki_clumps09,clumps_BR_13,2020MNRAS.497..648C,2023A&A...669A..21K,2023ApJ...948...77Y}. 
Density clumps and corresponding magnetic inhomogeneities  can significantly 
influence the energy distribution of PeV-regime particles, leading to a hardening 
of  their spectrum. Accounting for the clumps would require much more  
computational resources than the current model consumes, so we will leave this 
for a dedicated study of the issue. In a model way, this issue was 
addressed in Monte Carlo simulation by \citet{2032our}. They examined the effect of 
scattering  of cosmic rays by magnetic inhomogeneities (clumps in the wind of a massive star) in colliding winds 
in the gamma-ray binary PSR J2032+4127/MT91\:213 and showed that it can lead to 
a hard spectrum of PeV particles and  maximum energies of accelerated protons above 10 PeV. 
Monte Carlo modeling assumed a parameterization of particle scattering rate. Here by means 
of direct simulation of particle trajectories in magnetohydrodynamic flow produced by 
the colliding winds we showed that PeV-regime particles can be accelerated even in 
a simplified system modeled as an asymptotically homogeneous magnetized stellar wind flow 
colliding with a pulsar wind.

The acceleration of protons above 10 PeV,  which we observed in our simulations,
is of particular interest because such particles can exceed the threshold of 
the photo-meson reaction \citep[see e.g.][]{2004vhec.book.....A,2009herb.book.....D} 
on seed photons of optical/ultraviolet radiation from a bright young massive star. 
This process could allow gamma-ray binaries to produce  neutrinos and photons 
with energies above 100 TeV \citep[see e.g.][]{prediction-nu-Neronov2009,2032our}.
Recently, a $\sim 150$ TeV neutrino event associated with gamma-ray photons 
with energies above 300 TeV have been reported. Both signals originated from the 
Cygnus region containing PSR J2032+4127/MT91 213 and Cyg X3 \citep{Carpet2},
suggesting a multi-PeV proton acceleration regime 
if the signals are associated with the gamma-ray binary system \citep{2032our}.

There is evidence for the  possible  presence of a component 
of galactic origin among detected high-energy neutrino events
\citep[see e.g.][]{neronov_gal_neutr16,troitsky_neutrinos21,kovalev_neutrinos22}. 
Note that accompanying $\gsim$ 300 TeV gamma-rays cannot be detected from 
far away extragalactic sources \citep[see][]{Nikishov1962,2004vhec.book.....A}. 
The sub-population of gamma-ray binaries discussed above could contribute 
to the production of  PeV-regime cosmic rays and be responsible for some 
of the neutrino events in the Milky Way. 

There is about a dozen of confirmed gamma-ray binaries and in some of these the compact companion is a pulsar as we assumed above. 
Gauss-range magnetic field that is needed to accelerate PeV CRs, seems to be present in the winds interaction region in sources with the orbital periods shorter than a year.
The binary sources with much longer orbital periods like the recently discovered \psr2032 can accelerate PeV particles during a short part of the period near the periastron. On the other hand the number of such sources can be larger than we currently know because their TeV appearance is not very frequent. 
The observed luminosity of TeV radiation in \psr2032 around the orbit periastron is about 10$^{33} \ergs$ \citep{2018ApJ...867L..19A,2022icrc.confE.799K}. 
Models of the TeV radiation of gamma-ray binaries \citep[see e.g.][]{{2019A&A...627A..87C},takata17} considered multi-TeV leptons as the main source of the observed TeV regime radiation. The efficiency of the gamma-ray radiation due to hadronic collisions in the close vicinity of the wind collision region is not necessarily very high allowing the power of the accelerated particles production to be much higher than the detected gamma-ray emission. It should be noted, however that the photo-hadronic radiation process at very high energies can be very efficient \citep[see e.g.][]{2004vhec.book.....A,prediction-nu-Neronov2009,2032our}. While the observational constrains on the hadronic components in this systems are not yet available, nevertheless a  simplified Monte Carlo modeling \citet{2032our} which assumed some TeV proton injection rate into the wind colliding flows provided a high PeV proton acceleration efficiency. A non-negligible fraction of the plasma flows power can be transferred to PeV regime protons. Therefore, from the rough estimations the production power of the PeV regime CRs by the dozen of known gamma-ray binaries may reach $\sim$ 10$^{36} \ergs$.    
   
The power of the CR sources needed to maintain the directly observed fluxes of CR nuclei at the Earth orbit in the PeV regime estimated by \citet{Murase_Fukugita19,pevatron23} is $\sim$ 10$^{38} \ergs$ at 1 PeV. They got this CR production rate within a quasi-steady galactic CR propagation model. The estimation was obtained assuming that: 
   
   (i) the CRs of energies above PeV are distributed within the Galaxy in the same way as that of GeV-TeV CRs and 
   
   (ii) the confinement time of the PeV regime CRs can be obtained from the extrapolations of the traversed grammage of GeV-TeV nuclei which is derived from the detailed measurements of their isotopic composition.

While the gamma-ray maps and photon spectra obtained from the Milky Way with Fermi LAT detector are consistent with rather smooth CRs distribution over the Galaxy it remains to be proved with very high energy gamma-ray telescopes that the same is true for PeV regime CRs and their source distribution. At the moment PeV gamma-astronomy is the most perspective way to search the pevatrons.
The  gamma-ray binaries with high
magnetic field detected by the current generation of TeV-PeV regime Cherenkov telescopes may be among a few sources that can be detected by 
the dedicated PeV gamma-ray observatories -- 
LHAASO \citep[][]{LHAASONat21}, TAIGA \citep[][]{Taiga22} and  the others. 
Galactic Pevatrons --  sources of cosmic rays in the PeV-range -- are still being 
sought among objects of various types \citep[see e.g.][]{2019NatAs...3..561A,2020SSRv..216...42B,2021JPlPh..87a8401A,pevatron23}, and gamma-ray and neutrinos observation will help to estimate the relative role of the different type PeV CRs sources.

\section{Summary}
The main goal of our work is to study the ability of  gamma-ray binaries 
to accelerate multi-PeV energy cosmic rays, which may be responsible for 
gamma-ray and possibly high-energy neutrino emission of these systems. 
We injected  protons which are
assumed  to be pre-accelerated  to energies below the pulsar's potential drop  
(either by a rotation-powered pulsar 
or by some other process in the stellar winds) into time dependent domain with the colliding wind flows of highly magnetized plasma in the binary simulated with 2D RMHD model.  
Then we followed  the trajectories of the injected particles by  
directly  integrating their equations of motion (with 3D momenta) 
to study particle confinement and acceleration in a time-dependent system of the colliding
MHD flows. 
The acceleration rate of PeV particles strongly depends  on 
the magnitude of the magnetic field carried by the wind of the massive young star.
If this magnitude is in the Gauss-range, then binaries hosting a pulsar with 
a spin-down luminosity above 10$^{35}$ erg s$^{-1}$ can accelerate 
protons to PeV energy regime (well above the potential drop energy of an isolated  fast rotating pulsar) and radiate very high energy photons and neutrinos. 
The contribution of this type of pevatrons to the observed galactic cosmic ray fluxes depends on 
the number of the gamma-ray binaries which is not yet certain having in mind the presence of long 
orbital period binaries like the recently discovered \psr2032. However, this object can be considered as a candidate source of PeV regime radiation from the Cygnus region reported by LHAASO \citep{LHAASONat21} and Carpet 2 \citep{Carpet2}. Another perspective objects for a search of the very high energy radiation are  \lsi and LS 5039. The nature of a compact companion in gamma-ray binary LS 5039 is not yet established, but the presence of hard X-rays - soft gamma rays  synchrotron component with a very hard spectral index \citep{falanga21} may indicate the presence of colliding winds in the short period gamma-ray binary.       

\section{Acknowledgments}
\noindent  
The authors thank the two referees for constructive and useful comments. This research made use of PLUTO public RMHD code developed by A. Mignone and the PLUTO team. We acknowledge the use of  data provided by NASA ADS system and SIMBAD database, operated at CDS, Strasbourg, France. Simulations were performed partly at the Joint Supercomputer Center JSCC of the Russian Academy of Sciences, and partly at the \textsl{Tornado} subsystem of the Supercomputing Center at
the Peter the Great Saint-Petersburg Polytechnic University. Cosmic ray particle  spectra simulations  by  A.M.B.\ and A.E.P.\  were supported by the RSF grant 21-72-20020. Pulsar wind nebula model setup was made by K.P.L.\ and G.A.P.\ who were supported by the baseline project at the Ioffe Institute.

\bibliographystyle{model5-names}
\biboptions{authoryear}

\section*{Appendix A. RMHD model}
The acceleration of PeV-range protons in gamma-ray binaries is modeled 
using the results of a  2D relativistic MHD simulation 
of  the time-dependent structure of the wind collision region in these objects.
The simulation is based on the  \textsl{Relativistic MHD module}\ and 
the \textsl{Cosmic Ray particle module} of the code PLUTO  \citep{Mignone+07,Mignone18}.

The simulation is carried out in the reference frame of the pulsar.
At the beginning of each numerical run, we initiate a stationary uniform 
stellar wind flow with co-directional magnetic field and velocity, and fixed 
pressure and mass density. The flow is initiated throughout the entire 
computational domain, with the exception of a small area with radius 
$r_{in} = 0.15$ AU around a pulsar. From this area  the pulsar wind is 
injected. The pulsar wind model is described in Appendix B.

We employ 2D Cartesian grid evenly spaced in both dimensions. 
The parameters of the computational grids 
are given in  Table \ref{T1}. 
 The smallest scales ($\sim 0.01$--$0.1$ AU) allowed by the grid are 
$10^3$--$10^4$ times larger than the radii of the light cylinder of pulsars 
in gamma-ray binary systems. Yet, in a small vicinity of the polar axis 
of a pulsar wind nebula, the adopted resolution may still be  
insufficient for the cusps of the pulsar wind  termination shock 
to be fully detached from the area of pulsar wind injection.
This may have a minor effect on the plasma dynamics in the vicinity of the shock funnels 
(interfering with the formation of jets in the nebula), but is unlikely to affect the structure 
of nebular flows at the scales of interest.
In all runs, the boundary conditions  of free outflow for a background plasma and free escape for particles are adopted.

The inclination of the pulsar's rotation axis to the 
magnetic axis is described by the angle $\alpha$, and to the direction of 
the stellar wind magnetic field -- by the angle $\psi$.
In the present study we take $\psi = 45^{\circ}$ and $\alpha=45^{\circ}$. 
The pulsar's \textsl{inclination} $\alpha$  and spin-down luminosity $L$, and the pulsar wind's 
initial magnetization $\sigma_0$ and  Lorentz-factor $\varGamma_w$ are free parameters 
of the pulsar wind model (see Appendix B). Together with the stellar wind parameters they are listed in 
Table \ref{T1}. We consider a wide range of the massive star's magnetic fields and mass loss rates, 
and chose some reference values for the stellar wind  magnetic field and number density
(as described in Sec.\ \ref{M}).  
Some parameters whose values are considered fixed in all runs are given in the legend  of Table \ref{T1}.

For simplicity, we used the equation of state of an ideal plasma for the SW and neglected the radiative losses in simulation of local vicinities of colliding winds region. We also adopted a constant $\gamma = $ 4/3 for the both winds, following a typical 
approach \citep[see e.g., p.1.2.1 of][]{DelZannaOlmi17}
\begin{table}
\centering 
\resizebox{\columnwidth}{!}{%
\begin{tabular}{lcccccccc}\hline
    run & $B_{sw}$, G & $\psi, {}^{\circ}$ & L, erg/s & $\sigma_0$ & $\varGamma_w$ & box, au$^2$ & $N_y \times N_y$ & $\Delta$, au \\ \hline 
  A & 0.1 & 45 & $10^{37}$ & 3 & 100 & $40\times40$ & $1600 \times 1600$ & 0.025\\ 
    B & 1 & 45 & $10^{37}$ & 3 & 100 & $40\times40$ & $1600 \times 1600$ & 0.025\\ 
    C & 2 & 45 & $10^{37}$ & 3 & 100 & $40\times40$ & $1600 \times 1600$ & 0.025\\ 
    D & 3 & 45 & $10^{37}$ & 3 & 100 & $40\times40$ & $1600 \times 1600$ & 0.025\\ 
    E & 0.25 & 45  & $10^{37}$ & 0.3 & 100 & $40\times40$ & $1600 \times 1600$ & 0.025\\ 
    F & 0.5 & 45 & $10^{37}$ & 0.3 & 100 & $40\times40$ & $1600 \times 1600$ & 0.025\\ 
    G & 1 & 45 & $10^{37}$ & 0.3 & 100 & $40\times40$ & $1600 \times 1600$ & 0.025\\ 
    H & 2 & 45 & $10^{37}$ & 0.3 & 100 & $40\times40$ & $1600 \times 1600$ & 0.025\\ 
    I & 3 & 45 & $10^{37}$ & 0.3 & 100 & $40\times40$ & $1600 \times 1600$ & 0.025\\ \hline
\end{tabular}
}
\caption{Numerical RMHD models of a gamma-ray binary system.\\[.3ex]
\textsl{Computation domain:} \\
   \hspace*{0.05\columnwidth}\parbox[t]{.9\columnwidth}{%
  ``box'' -- size, \\
   $N_x$, $N_y$ --  number of grid cells in the $x$,\:$y$ directions\\
   $\varDelta = \varDelta_x = \varDelta_y$ --  numerical resolution of a uniform Cartesian  2D grid.
   } \\[.5ex]
\textsl{Parameters of the pulsar wind model} (see in Appendix):\ \\
    \hspace*{0.05\columnwidth}\parbox[t]{.9\columnwidth}{%
    $\varGamma_{w}$ -- the Lorentz factor,\\
    $L$ --  spin-down luminosity, \\ 
    $\sigma_0$ -- initial magnetization, \\
    $\alpha = 45^{\circ}$ -- angle between the rotational and magnetic axes of the pulsar. }\\[.5ex]
\textsl{Parameters of the stellar wind model}:\ \\
    \hspace*{0.05\columnwidth}\parbox[t]{.9\columnwidth}{%
    $B_{sw}$ -- magnetic field,\\
    $v_{sw} = 300\:$km$\cdot$s$^{-1}$ -- velocity, \\
    $p_{sw} = 10^{-7}\:$dyn$\cdot$cm$^{-2}$ -- pressure, \\
    $n_{sw} = 3 \cdot 10^4 \:$cm$^{-3}$ --  proton number density , \\
    $\psi$  -- angle between the pulsar’s rotational axis and $\bm{B}_{sw}$.} \\[.5ex]
\textsl{Notes:}\;\; An estimate of $n_{sw}$ is given  for the distance (from a Be-star) of $10^{14}\:$cm
    \hspace*{0.05\columnwidth}\parbox[t]{.95\columnwidth}{%
    and the stellar mass loss rate $\dot{M}_{-7} \sim 0.01 \:\Msun\cdot$yr$^{-1}\:$.
    An estimate of  $v_{sw}$ refers for  a Be-star equatorial outflow velocity 
    as seen in the pulsar's rest frame.
    A grid resolution $\varDelta$ is suitable for studying the confinement and acceleration
    of very high energy particles of PeV regime.
    }
}
\label{T1}
\end{table}

\section*{Appendix B. Pulsar wind model}

We applied a pulsar wind model of \citet{Porth+14}.
The free parameters of the model are: the spin-down luminosity $L$ and magnetic inclination 
$\alpha$ of the pulsar,  and the  Lorentz factor $\varGamma_w$ and initial magnetization 
$\sigma_0$ of the pulsar wind. Their values  are listed in Table \ref{T1}.

The model assumes a latitudinally anisotropic wind with an energy flux density 
depending on colatitude $\theta$ and radial coordinate $r$ as
\begin{equation}
    f_{tot\, }(r, \theta) = \frac{L}{L_{\, 0}} \frac{1}{r^{\, 2}}\cdot 
    \left( \sin^2 \theta + \varepsilon  \rule{0pt}{1.5ex} \right)\, .
\end{equation}
Here $L$ is a pulsar spin-down luminosity, $L_{\,0} = 4\pi\:(2/3 + \varepsilon)$ is a normalization
constant, and $\varepsilon = 0.02$ prevents the energy flux from vanishing at the poles.
The wind is cold, so its total  energy flux is divided into magnetic and kinetic components,
$f_{tot} = f_m + f_k$:
\begin{equation}
    f_{m\,}(r, \theta) = \frac{\sigma(\theta)\cdot f_{tot\:}(r, \theta)}{1 + \sigma(\theta)}; \quad
    f_{k\,}(r, \theta) = \frac{f_{tot\:}(r, \theta)}{1 + \sigma(\theta)} \: .
    \label{eq:fluxes}
\end{equation}
The ratio $f_m/f_k$ determines  the wind  magnetization $\sigma$ 
(upstream of the termination shock) at a given colatitude $\theta$ :
\begin{equation}
    \sigma(\theta) = \frac{\tilde{\sigma}(\theta)\cdot \chi_{\alpha} (\theta)}{1 + \tilde{\sigma}(\theta)\cdot 
    (1 - \chi_{\alpha} (\theta)\:)} \: .
    \label{eq:sigma}
\end{equation}
The magnetic field of the pulsar wind is  frozen into the plasma and purely toroidal. 
The toroidal field must vanish at the poles and change sign at the rotational equator 
of the pulsar.  Eq.\ \ref{eq:sigma} accounts for these features with the functions
$\tilde{\sigma}$ and $\chi$, respectively:
\begin{eqnarray}
    \tilde{\sigma}(\theta) = \sigma_0 \cdot {\rm min} \left\{ \:1\, ,\;\; 
    \theta^2/\theta_0^2\, , 
    \;\;     (\pi - \theta)^2/\theta_0^2  \rule{0pt}{1.ex} \;\;\right\} \: ;  \label{eq:sigma-tilde}\\
    \chi_{\alpha}(\theta) = 
    \begin{cases} 
    (2 \phi_a (\theta)/\pi - 1)^2 & \mbox{if } |\pi/2 - \theta| < \alpha  \\
        1 & \mbox{otherwise}
    \end{cases} \: , \label{eq:chi}
\end{eqnarray}
with $\phi_{\alpha} (\theta) = \arccos{(-\cot(\theta) \cot(\alpha))}\:$.
Here $\theta_0 = 10^{\circ}$, and $\sigma_0$ is an initial magnetization 
of the wind at the light cylinder of the pulsar. 
As a result of the sign change, the wind of a rotating pulsar with a magnetic 
inclination $\alpha$ carries stripes of alternating magnetic polarity, filling 
the equatorial sector of the wind with an angular extent of $\pm \alpha$. 
The stripes are supposed to annihilate, either on the way to the termination shock,
or directly on it \citep{Lyubarsky+03, Komissarov13, Cerutti+20}. 
Hence, within the striped wind sector subtended by an angle
$2\alpha$, the magnetic field should dissipate completely, as described by 
Eqs.\ \ref{eq:sigma}--\ref{eq:chi}. In the simulation, we take $\alpha=45^\circ$.
The magnetic field and density 
are then calculated using Eq.\ (\ref{eq:fluxes}):
\begin{equation}
    B_{pw\,}(r, \theta) = \pm \sqrt{\:4 \pi f_{m\, } (r, \theta) / c \rule{0pt}{1.8ex}}\: , 
    \quad
    \rho_{pw\,}(r, \theta) = \frac{f_{k\,} (r, \theta)}{\varGamma_w^{\,2} \,c^3}\: .
\end{equation}
The wind velocity is directed radially outwards; its magnitude in terms of the 
Lorentz-factor $\varGamma_w$ is 
\begin{equation}
    u_{pw}/c  = \sqrt{\:1 - \varGamma_w^{\, -2}\:}.
\end{equation} 
On the inner boundary of the wind, pressure
is calculated using the equation of state  
for an ideal relativistic plasma (with the adiabatic index $\gamma = 4/3$).
Radial and angular coordinates are expressed  in terms of Cartesian coordinates as
\begin{equation}
r = \sqrt{\:x^2 + y^2}\; ;  \quad \theta = \arccos\, \left( \,y\,/\!\sqrt{x^2 + y^2}\, \right)   
\end{equation}

\section*{Appendix C. Test particle propagation}

Propagation and acceleration of PeV-regime protons are simulated using the particle 
moving unit of the PLUTO code (\textsl{Cosmic Ray}\ module; \citealp{Mignone18}). 
In this module the MHD-PIC model is implemented, and accelerating particles are 
treated with the particle-in-cell techniques. The latter allow one to resolve 
 the particle gyration (Larmor) scales. In the setups $\mbox{A}$--$\mbox{D}$, 
 10,000 protons  with an initial energy of 2.4 PeV are 
injected near the PWN boundary  1/6 day after the start of the PWN evolution.
At this point, the nebula is already fully developed.
After injection, protons are treated as test-particles that do not affect the dynamics 
of the background plasma. Their propagation is followed simultaneously with the flow 
dynamics using the MHD-PIC formalism implemented in PLUTO.
To study the spectral evolution of the accelerated particles, in the setups 
$\mbox{E}$--$\mbox{I}$ we inject 160,000 particles with an energy of 0.1 PeV.
\end{document}